\documentclass[letterpaper,notitlepage,nofootinbib,superscriptaddress]{revtex4-1} 

\usepackage[utf8]{inputenc}
\usepackage{amsmath,amssymb,amsfonts}
\usepackage{graphicx,wrapfig}
\usepackage{dcolumn}
\usepackage{bm}
\usepackage{color}
\usepackage{bbold}
\usepackage{url}
\usepackage{slashed}
\usepackage[svgnames]{xcolor}
\usepackage[english]{babel}
\usepackage{microtype}

\usepackage[colorlinks=true,linkcolor=blue,urlcolor=blue,citecolor=blue]{hyperref}

\usepackage{color}

\definecolor{darkgreen}{rgb}{0,0.65,0}

\newcommand{\be}{\begin{eqnarray}}
\newcommand{\ee}{\end{eqnarray}}
\newcommand{\ba}{\begin{array}}
\newcommand{\ea}{\end{array}}

\begin{document}

\title{${\bf \Omega(2012)}$ through the looking glass of flavour ${\bf SU(3)}$ }

\author{Maxim V.~Polyakov}
	\affiliation{Petersburg Nuclear Physics Institute, 
		Gatchina, 188300, St.~Petersburg, Russia}
	\affiliation{Institut f\"ur Theoretische Physik II, 
		Ruhr-Universit\"at Bochum, D-44780 Bochum, Germany}
\author{Hyeon-Dong Son}
	\affiliation{Institut f\"ur Theoretische Physik II, 
		Ruhr-Universit\"at Bochum, D-44780 Bochum, Germany}
		\author{Bao-Dong Sun }
	\affiliation{Institute of High Energy Physics, Chinese Academy of Sciences, Beijing 100049, P. R. China}
	\affiliation{School of Physics, University of Chinese Academy of Sciences, Beijing 100049, P. R. China}
\author{Asli Tandogan}
	\affiliation{Institut f\"ur Theoretische Physik II, 
		Ruhr-Universit\"at Bochum, D-44780 Bochum, Germany}

\begin{abstract} 
We perform the flavour $SU(3)$ analysis of the recently discovered $\Omega(2012)$ hyperon. We find that well known (four star) $\Delta(1700)$ resonance with quantum numbers of $J^P=3/2^-$
is a good candidate for the decuplet partner of $\Omega(2012)$ if the branching for the three-body decays of the latter is not too large $\le 70$\%. 
That implies that the quantum numbers of $\Omega(2012)$ are  $I(J^P)=0(3/2^-)$. The predictions for the properties of still missing
$\Sigma$ and $\Xi$ decuplet members are made. We also discuss the implications of the ${ \overline{ K} \Xi(1530)}$ molecular picture of $\Omega(2012)$. 
  Crucial experimental tests to distinguish various pictures of $\Omega(2012)$ are suggested.
\end{abstract}


\maketitle

\section*{\normalsize \bf Introduction}
\begin{figure}[h!]
\centering
\includegraphics[width=7.8 cm]{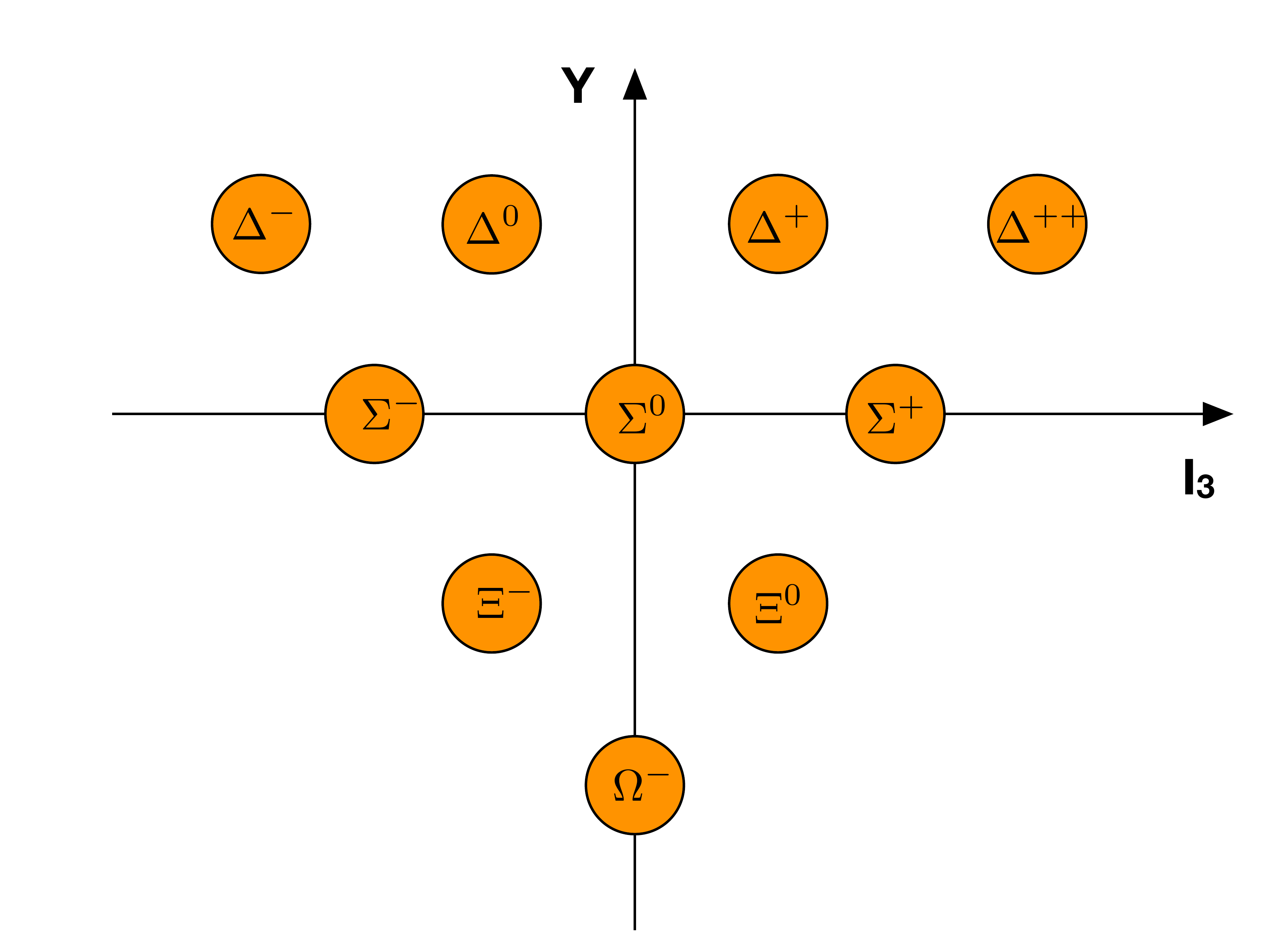}
\vspace{-10pt}
\caption{$SU(3)$ decuplet.}
\label{Fig:decuplet}
\end{figure}

An $\Omega$ hyperon  is a baryon with strangeness $S=-3$. It is a good marker of the flavour $SU(3)$ decuplet, if its isospin is zero, or of the exotic $27$-plet, if its isospin is one. The $\Delta$ partner is another unique marker for the decuplet. 
If the $\Omega$ is in the decuplet then it must be accompanied by the $\Xi, \Sigma$ and $\Delta$ flavour partners, see Fig.~\ref{Fig:decuplet}.

\noindent
 Moreover, if we know the masses of $\Delta$ and $\Omega$ from  the decuplet, we can predict masses
of the $\Sigma$ and $\Xi$ partner using the approximate $SU(3)$ symmetry. Additionally, the $SU(3)$ symmetry allows to predict the partial decay widths of all partners if we know one of decuplet partners, e.g. decays of $\Omega$.

The recently discovered $\Omega(2012)$ hyperon by the BELLE collaboration has a mass and width \cite{Yelton:2018mag} :

\be
\label{Eq:masswidthexp}
M=2012.4 \pm 0.7 {\rm (stat)} \pm 0.6 {\rm (syst)}{\rm MeV},\,\,   \Gamma_{\rm tot}=6.4^{+2.5}_{-2.0} {\rm (stat)} \pm 1.6{\rm  (syst)}{\rm MeV},
\ee
neither the spin-parity nor the isospin are  measured.

Our aim here is to identify the $\Delta$ partner for the recently discovered $\Omega(2012)$. The information about the properties of the $\Delta$ resonance, e.g. mass, partial widths and quantum numbers,
are obtained e.g. from well studied $\pi N$ scattering. Therefore, using the approximate flavour $SU(3)$ symmetry we can check which one of the known $\Delta$ resonances belongs to the same
decuplet as $\Omega(2012)$. Finding the $\Delta$ partner, we can establish the quantum numbers of $\Omega(2012)$ and make predictions for the properties of the other, $\Sigma$ and $\Xi$, partners of the corresponding decuplet. Additionally, as the properties of the new $\Omega$ hyperon are well measured, we can narrow down the uncertainties in the properties of its $\Delta$ partner.

Another possible interpretation of the $\Omega(2012)$ is the  ${ \overline{ K} \Xi(1530)}$  bound state. In this molecular picture the isospin of $\Omega(2012)$ can
be either zero (decupet) or one (27-plet). We discuss below the predictions for such picture.

\noindent
\section*{\normalsize \bf Identification of the ${\bf \Delta}$ decuplet partner for ${\bf \Omega(2012)}$}

To make an identification of the $\Delta$ partner of $\Omega(2012)$ we use the flavour $SU(3)$ formalism described in details in reviews \cite{Samios:1974tw,Guzey:2005vz}.
According to Gell-Mann--Okubo mass formula, the mass splitting in the decuplet is equidistant with typical mass splitting of order $100$-$150$~MeV. Therefore, we can expect the
$\Delta$ partner of $\Omega(2012)$ in the mass range of 1500-1700~MeV. In this mass range, Particle Data Group (PDG) \cite{Patrignani:2016xqp} reports three $\Delta$ resonances;
the $\Delta(1600)$ with $J^P=3/2^+$, the $\Delta(1620)$ with $J^P=1/2^-$, and the $\Delta(1700)$ with $J^P=3/2^-$.
In the most recent global $SU(3)$ analysis, \cite{Guzey:2005vz}, the $\Xi$ and $\Omega$ decuplet partners of these  $\Delta$'s are missing\footnote{The $\Sigma$ partner was clearly identified in Ref.~\cite{Guzey:2005vz} only for $\Delta(1620)$, it is $\Sigma(1750)$ with $J^P=1/2^-$. }.
For the $\Omega(2012)$, we can repeat the analysis to see which decuplet it may belong to.

 There are only three open strong decay channels, $\Xi \overline{ K}$, $\Xi \overline{ K} \pi$ and $\Omega^-\pi\pi$\footnote{For the isovector
$\Omega(2012)$ from the 27-plet an additional strong decay $\pi\Omega^-$ channel opens.} for the $\Omega(2012)$ from the decuplet,, 
so we can use the sum of three-body decay branching ratios:
 
\be
\label{Eq:brr}
b=1-\frac{\Gamma\left(\Omega(2012)\to \Xi \overline{ K}  \right)}{\Gamma_{\rm tot}}, \ \ 0\le b\le 1
\ee
as a single free parameter and calculate the two particle decay of $\Omega(2012)$ as
\be
\Gamma(\Omega(2012)\to \Xi \overline{ K}) =(1-b)\cdot \Gamma_{\rm tot},
\ee
where the experimental total width is given by Eq.~(\ref{Eq:masswidthexp}). 
By knowing the partial decay width of $\Omega(2012)$ we can now make predictions for the partial decay
widths of $\Delta$ decuplet partner and compare them with the experimental data on the $\Delta$ resonances depending on the free parameter $b$.

For the partial decay widths of the decay $B_1\to B_2 M$ ($M$ is a pseudoscalar meson of the mass $m$) we shall use two different formulae which differ by $SU(3)$ symmetry-violating corrections of order $O(m_s)$. 
In this way we can estimate the systematic uncertainties of our $SU(3)$ analysis.
The first formula used in the global $SU(3)$ analyses of Refs.~\cite{Samios:1974tw,Guzey:2005vz} is:
\be
\label{Eq:formula1}
\Gamma\left( B_1\to B_2 M\right)=\frac{g_{B_1B_2M}^2}{8 \pi}\left( \frac{k}{M_0}\right)^{2 J+ {\cal N}} \left( \frac{k}{M_1}\right) M_0.
\ee
The second formula used was obtained in Ref.~\cite{SemenovTianShansky:2007hv}. It is in the framework of effective chiral Lagrangian for baryon resonances of any spin, and defined as:
\be
\label{Eq:formula2}
\Gamma\left( B_1\to B_2 M\right)=\frac{g_{B_1B_2M}^2}{8 \pi}\left( \frac{k}{M_0}\right)^{2 J-1} \left( \frac{k}{M_1}\right)  \frac{[(M_1-{\cal N} M_2)^2-m^2]}{M_1}.
\ee
In both formulae $J$ is the spin of decaying baryon, ${\cal N}=P\left(-1\right)^{J-1/2}$ is its {\it normality} \cite{Carruthers:1966kee,SemenovTianShansky:2007hv}, $P$ is the parity of the resonance.
$M_0$ is the mass constant parameter characterising details of the baryon dymanics\footnote{For example, $M_0=f_\pi$ in the context of the effective chiral Lagrangian
approach of  ef.~\cite{SemenovTianShansky:2007hv}.}, its precise value is irrelevant for our analysis  and we, following \cite{Samios:1974tw}, choose $M_0=1$~GeV. Eventually  $k$ is the c.m.s momentum 
\be
k=\frac{\sqrt{(M_1^2-(M_2+m)^2) (M_1^2-(M_2-m)^2)}}{2 M_1}.
\ee
 The coupling constants $g_{B_1B_2M}$ for the various decays ${\bf 10\to 8+8}$ are related to each other by  $SU(3)$ Clebsch-Gordan coefficients 
listed in Table~\ref{Tab:su3}.

\begin{table}[h!]
\centering
	\begin{tabular}{  l | l | r| r}
    \hline\hline
    \textbf{10}&Decay Mode &$\rightarrow$ \textbf{8+8}  & $\rightarrow$\textbf{10+8} \\ \hline
      & &&  \\
    $\Delta$ & $\rightarrow$ N$\pi$& -$\sqrt{2}/2$ $A_{10}$ &    \\ 
      & $\rightarrow$ $\Sigma$ K& $\sqrt{2}/2$ $A_{10}$ &  \\ 
      & $\rightarrow$ $\Delta$ $\pi$ & &$\sqrt{10}/4$ $A_{10}^\prime$   \\ 
       & $\rightarrow$ $\Delta$ $\eta$ & &$-\sqrt{2}/4$ $A_{10}^\prime$   \\ 
       & $\rightarrow$ $\Sigma^*$ K & &$1/2$ $A_{10}^\prime$   \\ 
        & &&  \\ \hline
        & & & \\
      $\Sigma$&$\rightarrow$ $\Lambda$ $\pi$&$- 1/2 A_{10}$& \\
       & $\rightarrow$ $\Sigma$ $\pi$,\, $\Xi$ K &$\sqrt{6}/6 A_{10}$& \\
        &  $\rightarrow$ N $\bar{K}$&-$\sqrt{6}/6 A_{10}$& \\ 
         & $\rightarrow$ $\Sigma$ $\eta$ &$1/2 A_{10}$&  \\ 
         &$\rightarrow$ $\Sigma^*$ $\pi$,\, $\Xi^*$ K,\, $\Delta \bar{K}$ & &$\sqrt{3}/6 A_{10}^\prime$\\
         &$\rightarrow$ $\Sigma^*$ $\eta$ & &0\\
          & &&  \\ \hline
           & & & \\
    $\Xi$&$\rightarrow$ $\Xi$ $\pi$,\, $\Xi \eta$,\, $\Sigma \bar{K}$&$1/2 A_{10}$& \\
           & $\rightarrow$ $\Lambda$ $\bar{K}$ &$-1/2 A_{10}$& \\
             & $\rightarrow$ $\Xi^*$ $\pi$,\, $\Xi^* \eta$ & &$\sqrt{2}/4 A_{10}^\prime$ \\ 
             & $\rightarrow$ $\Sigma^*\bar{K}$ & &$\sqrt{2}/2 A_{10}^\prime$ \\ 
             & $\rightarrow$ $\Omega$ K & &$1/2 A_{10}^\prime$ \\  
              & &&  \\\hline
             & & & \\
     $\Omega$&$\rightarrow$ $\Xi$ $\bar{K}$&$A_{10}$& \\  
         &$\rightarrow$ $\Xi^* \bar{K}$,\, $\Omega \eta$& &$\sqrt{2}/2A_{10}^\prime$\\
          & &&  \\ 
         \hline\hline
  \end{tabular}
\caption{SU(3) relations for coupling constants $g_{B_1B_2M}$ for \textbf{10$\rightarrow$8+8} and \textbf{10$\rightarrow$\textbf 10+8}  transitions}.
\label{Tab:su3}
 \end{table}

\begin{figure}[h]
\centering
\includegraphics[width=15 cm]{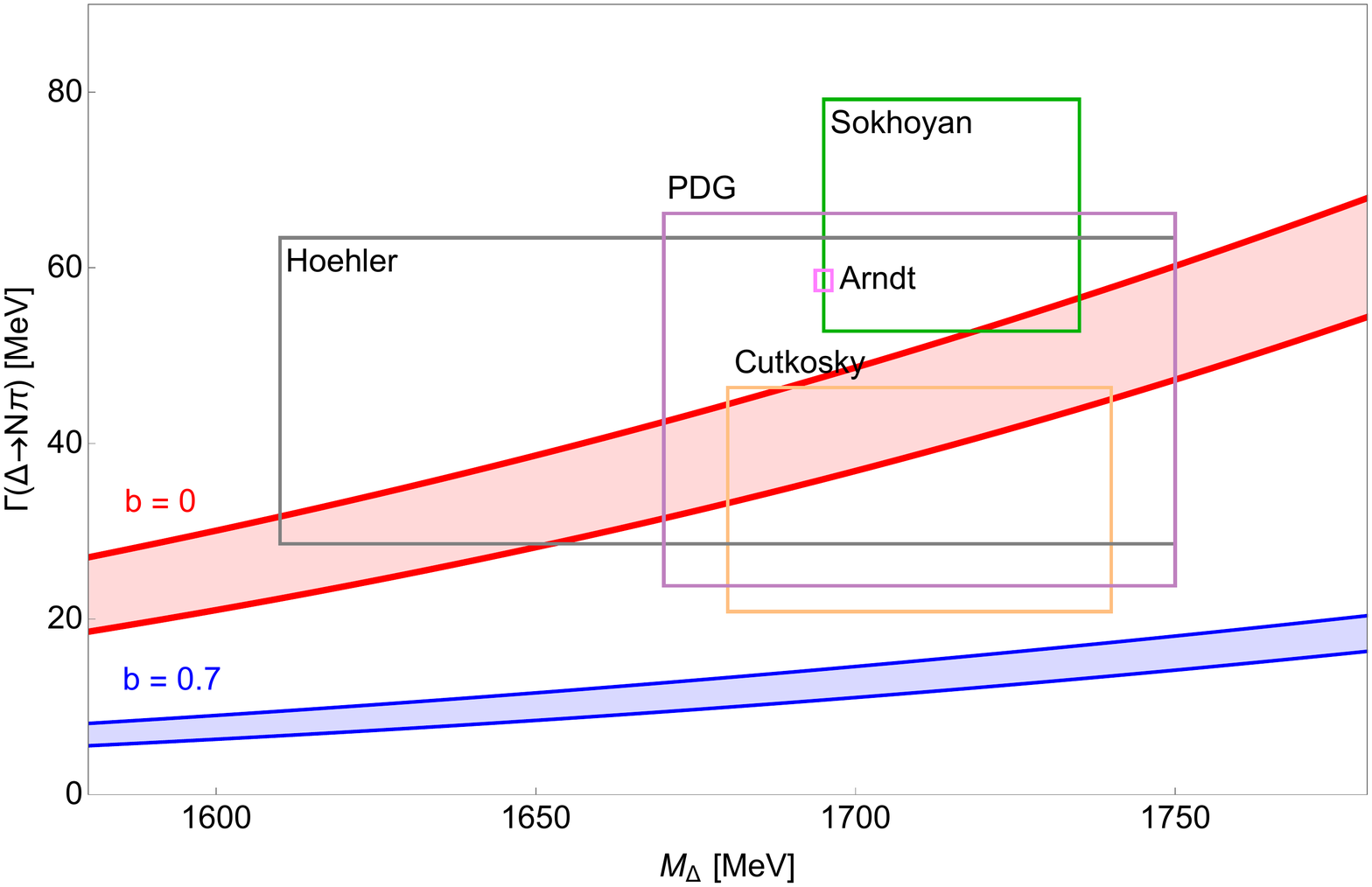}
\caption{$\Gamma(\Delta\rightarrow \pi N)$ versus $M_\Delta$. Red band is our prediction of the decay width $\Gamma(\Delta\rightarrow \pi N)$  based on Eq.~(\ref{Eq:formula1}) and Eq.~(\ref{Eq:formula2})
and the parameter $b=0$ Eq.~(\ref{Eq:brr}). Blue band corresponds to $  b=0.7$. The rectangles show the PWA analyses \cite{Cutkosky:1980rh,Forsyth:1980qy,Hohler:1979yr,Arndt:2006bf,Sokhoyan:2015fra} results and PDG estimations \cite{Patrignani:2016xqp}.}
\label{Fig:comparison}
\end{figure}

From Table~\ref{Tab:su3} we see that the partial decay ${\bf 10\to 8+8}$ widths for all members of the decuplet are fixed in terms of one constant $A_{10}$. This constant can be constrained
by the experimental values for mass and the width of $\Omega(2012)$ by assuming various quantum numbers of $\Omega(2012)$. By doing this we predict 
that the $\Delta(1600)$ with $J^P=3/2^+$ and the $\Delta(1620)$ with $J^P=1/2^-$ are excluded as decuplet partners of the $\Omega(2012)$ because, from $SU(3)$ relations, the resulting
partial decay widths of $\Delta$'s are much smaller than the ones listed in PDG \cite{Patrignani:2016xqp}, even for the value of parameter $b=0$.

 For $\Delta(1700)$ with $J^P=3/2^-$ we obtain that its $\pi N$ partial decay  width is in 
good agreement with the width from the well measured width of $\Omega(2012)$  if the parameter $b$ (see Eq.~(\ref{Eq:brr})) is not too large. 
This is illustrated on  Fig.~\ref{Fig:comparison} where we compare our $SU(3)$ prediction,
for the $\pi N$ partial decay width of $\Delta(1700)$, with results of various PWA \cite{Sokhoyan:2015fra, Arndt:2006bf, Cutkosky:1980rh, Forsyth:1980qy, Hohler:1979yr} used by PDG in their estimates of
the $\Delta(1700)$ properties. 

In Fig.~\ref{Fig:comparison}, we plot two bands for the predicted $\pi N$ partial width of $\Delta(1700)$ for two values of the parameter $b=0$ and $b=0.7$ by the  $SU(3)$  analysis.
The width of each band indicates our systematic uncertainties arising from the use of two different formulae for the partial decay width, Eq.~(\ref{Eq:formula1}) and Eq.~(\ref{Eq:formula2}),
as well as the statistical uncertainty due to experimental error bars in Eq.~(\ref{Eq:masswidthexp}).

Additionally our analysis allows to narrow down the $\pi N$ partial width and mass of the $\Delta(1700)$ giving a slight preference to analyses of Cutkosky et al. \cite{Cutkosky:1980rh} and Hoehler et al. \cite{Hohler:1979yr}, while disfavouring slightly
the PWA of Sokhoyan et al. \cite{Sokhoyan:2015fra} and Arndt et al. \cite{Arndt:2006bf}.

From Fig.~\ref{Fig:comparison}
we see that if the parameter $b\ge 0.7$ then the data on $\Delta(1700)$ decays are not compatible with its interpretation as the decuplet 
partner of $\Omega(2012)$. In this case, one of the interpretations of the $\Omega(2012)$ hyperon can be a ${ \overline{ K} \Xi(1530)}$ bound state with quantum numbers $J^P= 3/2^-$ ($S$-wave bound state).
In these two interpretations, we obtain that  the quantum numbers of the recently discovered 
$\Omega(2012)$ are $J^P=3/2^-$. 

The excited $\Omega$ hyperon with the mass around 2000~MeV and quantum numbers $I(J^P)=0(3/2^-)$ was obtained in different models before the discovery
of $\Omega(2012)$; in the quark model \cite{Faustov:2015eba},
in the lattice simulations \cite{Engel:2013ig} and in the Skyrme model \cite{Oh:2007cr}. {We note, however, that in the quark model calculations of Ref.~\cite{Faustov:2015eba},
the equidistant mass splitting between the decuplet partners is strongly violated. The lattice calculations of Ref.~\cite{Engel:2013ig} give rather large mass sptlitting in the discussed decuplet of 
$\delta_{10}=155\pm 50$~MeV, although compatible with the result of our analysis $\delta_{10}=104 \pm 15$~MeV within the error bars.}

After its discovery the $\Omega(2012)$ hyperon was considered in the framework of chiral quark model Ref.~\cite{Xiao:2018pwe} and in the framework of QCD sum rules in Ref.~\cite{Aliev:2018syi}. In both studies, the authors obtained
$J^P=3/2^-$ quantum numbers as well. The decay of  $\Omega(2012)$ in the chiral quark model is dominated by $\Xi \overline{ K}$ mode, so the ${ \overline{ K} \Xi(1530)}$ molecular scenario
is excluded in chiral quark model.  The molecular scenario is also excluded in the framework of the QCD sum rules of \cite{Aliev:2018syi} as the width of $\Omega(2012)$ is dominated by the $\Xi \overline{ K}$ decay mode  \cite{Aliev:2018yjo}.
Therefore, the predictions of chiral quark model of Ref.~\cite{Xiao:2018pwe}   and of the QCD sum rules \cite{Aliev:2018syi,Aliev:2018yjo} can be tested -- both approaches should predict $J^P=3/2^-$ $\Delta$ with the decay properties as shown on Fig.~\ref{Fig:comparison}.
Unfortunately, $SU(3)$ partners of the $\Omega(2012)$ hyperon were not discussed in Refs.~\cite{Xiao:2018pwe,Aliev:2018syi,Aliev:2018yjo}. 

\noindent
\section*{\normalsize \bf Properties of new ${\bf J^P=3/2^-}$ decuplet of baryons}

We obtain that if the parameter $b$ of Eq.~(\ref{Eq:brr}) is smaller than 0.7,  the $\Omega(2012)$ hyperon and $\Delta(1700)$ are good candidates for  the members of the same decuplet.
The mass splitting $\delta_{10}$ in the decuplet is equidistant and is $\delta_{10}=104 \pm 15$~MeV, where the error bar is dominated by the uncertainty of the $\Delta(1700)$ mass. Using this value we can predict
properties of other members of the decuplet, i.e. $\Sigma$ and $\Xi$ hyperons. Their quantum numbers are $J^P=3/2^-$ and masses $M_\Sigma=1805\pm 40$~MeV and $M_\Xi=1910\pm 40$~MeV.

In principle, the masses of the $\Sigma$ and $\Xi$ decuplet members can be shifted from the equidistant rule due to the mixing with nearby members of $J^P=3/2^-$ octet. However, according
to the analyses of Refs.~\cite{Samios:1974tw,Guzey:2005vz} there are no such nearby states, so we expect the Gell-Mann--Okubo mass formula should work well for the new decuplet.

There are no candidates for suggested  $\Sigma$ and $\Xi$ decuplet states in PDG, so their existence is our prediction. Our predictions for the partial decay widths for $\Sigma$ and $\Xi$ members of the new decuplet are shown in Table~\ref{tab.:sigmaxi}. In this table we note that the $\Xi$ of the discussed decuplet can be identified with known $\Xi(1950)$. However, in Ref.~\cite{Guzey:2005vz} it was shown that the $\Xi(1950)$
is fitted very well into the octet of baryons with $J^P=5/2^-$.

\begin{table}[h!]
\centering
	\begin{tabular}{  l | l  c c }
    \hline\hline
      &Decay Mode &$\Gamma_{\mathrm{Eq.(\ref{Eq:formula1})}}$ [MeV] & $\Gamma_{\mathrm{Eq.(\ref{Eq:formula2})}}$ [MeV]\\ \hline
      & &  \\
      $\Delta$ & $\rightarrow$ $N\pi$          & 29 - 44 & 39 - 58   \\ 
               & $\rightarrow$ $\Sigma K$      & 0 - 0.2 		 & 0 - 2   \\
	           & &  \\ \hline 
               & &  \\
      $\Sigma$ & $\rightarrow \Lambda\pi$      & 10 - 16  & 11 - 18 \\
      		   & $\rightarrow \Sigma\pi$       & 4 - 7	 & 4 - 7   \\
       		   & $\rightarrow$ $N\bar{K}$      & 5 - 10   & 7 - 12  \\
       	       & $\rightarrow$ $\Sigma$ $\eta$ & 0.01- 0.5   & 0.01 - 0.5   \\
       		   & &  \\ \hline
       		   & &  \\
      $\Xi$    & $\rightarrow\Xi\pi$           &  5 - 9  & 5 - 9   \\
               & $\rightarrow\Sigma \bar{K}$   &  2 - 5  & 2 - 5   \\
               & $\rightarrow\Lambda \bar{K}$  &  5 - 9  & 5 - 10 \\
               & $\rightarrow\Xi \eta$         &  0 - 0.3  & 0 - 0.3   \\
               & &  \\ \hline
               & &  \\
      $\Omega$ & $\rightarrow\Xi\bar{K}$       & Input & Input \\

          & &  \\ \hline\hline
  \end{tabular}
\caption{Partial decay width predictions for the missing members of the new $J^P=3/2^-$ decuplet assuming $b=0$ (see Eq.~(\ref{Eq:brr})).
The predictions for other values of $b$ can be obtained by rescaling above numbers by $(1-b)$. 
Masses of these states are predicted in the range $M_\Sigma=1805\pm 40$~MeV and $M_\Xi=1910\pm 40$~MeV.
The larger value of decay widths correspond to the larger mass.}
\label{tab.:sigmaxi}
 \end{table}

In Table~\ref{tab.:sigmaxi} we gave predictions for the partial decay width using two different formulae Eq.~(\ref{Eq:formula1}) and Eq.~(\ref{Eq:formula2}) to illustrate the systematic uncertainties inherent
to our $SU(3)$ analysis.

\noindent
\section*{\normalsize \bf On ${\bf \Omega(2012)}$ as  a ${\bf \overline{ K} \Xi(1530)}$ molecule}

The mass of $\Omega(2012)$ is slightly below of the ${ \overline{ K} \Xi(1530)}$ threshold, so it could be the corresponding molecular state. If the ${ \overline{ K} \Xi(1530)}$ bound state is in $S$-wave the
quantum numbers of the molecula are $J^P=3/2^-$, its isospin can be either zero (decuplet) or one (27-plet).

 We stress that the flavour $SU(3)$ formalism (Gell-Mann--Okubo mass formulae, relations between widths, etc.) does not work for the molecular states -- the corresponding $SU(3)$ breaking corrections
are of order $\sim m_s/({\rm binding\ energy})$ (not $\sim m_s/({\rm hadron\ mass})$ as for genuine resonances) and hence can be very large. For example, the deuteron nucleus belongs to the $SU(3)$
anti-decuplet of $B=2$ hypernuclei, but the corrections to the Gell-Mann-Okubo formula are large enough such that one cannot make precise predictions for the corresponding partner hypernuclei.
The best one can do from the $SU(3)$ analysis is to identify the most attractive channels in hyperon-hyperon potential.

To obtain the $\Omega(2012)$ as the ${ \overline{ K} \Xi(1530)}$  bound state, one needs an attraction in the $S$-wave. 
Such an attraction can be provided by the Weinberg-Tomozawa (WT) term of the effective chiral Lagrangian. Detailed studies of that were performed in Refs.~\cite{Kolomeitsev:2003kt,Sarkar:2004jh}.
The authors of Ref.~\cite{Kolomeitsev:2003kt} found an isospin zero (decuplet) ${ \overline{ K} \Xi(1530)}$ bound  state with the mass of 1950~MeV. Probably, tuning the model parameters,
one can adjust the mass of the state to $\sim 2000$~MeV. However, the channel with isospin one (27-plet) was not studied quantitatively in Ref.~\cite{Kolomeitsev:2003kt}. 

The studies, very similar to \cite{Kolomeitsev:2003kt} in methods, but using different to \cite{Kolomeitsev:2003kt} subtractions, were performed in Ref.~\cite{Sarkar:2004jh}. The authors came to 
different results for the isospin zero channel with $S=-3$ (decuplet).
In \cite{Sarkar:2004jh}  it was obtained that the bound ${ \overline{ K} \Xi(1530)}$ state does not exists in this channel, instead the two coupled ${ \overline{ K} \Xi(1530)}$ and $\eta\Omega^-$
states with the pole position at $2141-i 38$~MeV were found. This pole is far away from $\Omega(2012)$. Again the studies of isospin one (27-plet) channel were not performed.
The acute difference of the  results for the $S=-3, I=0$ channel of Ref.~\cite{Kolomeitsev:2003kt} and Ref.~\cite{Sarkar:2004jh} calls for clarification. Also, potentially
very interesting $S=-3, I=1$ channel should be studied.

On general grounds we can estimate that,
if a ${ \overline{ K} \Xi(1530)} $ bound state is formed in the isospin zero channel (decuplet), its main decay mode is $\Xi \overline{ K} \pi$ with 
the corresponding partial width of $\sim 10$~MeV (order of the $\Xi(1530)$ width). {Recently the authors of \cite{Valderrama:2018bmv,Lin:2018nqd}
 confirmed, by model calculations in the effective field theory, our qualitative conclusion that the dominant decay mode of ${ \overline{ K} \Xi(1530)} $ bound state is
$\Xi \overline{ K} \pi$. However, they obtained for the corresponding partial decay width the value considerably smaller than our estimate of $\sim 10$~MeV. They attributed this
difference to  the binding energy of the molecule as well as to the kinetic energy of $K^-$  inside the molecule. We note, however, that the naive phase space reduction of the width due to binding energy 
frequently cancels with the final state interactions, see examples of such cancelations for muon atoms in Refs.~\cite{Uberall:1960zz,Czarnecki:1999yj}. 
It could be that in the case of the ${ \overline{ K} \Xi(1530)} $ bound state  similar cancelation
happens, we shall study this elsewhere.}

For the isospin one molecule  (27-plet) we expect larger width of $\sim 30-50$~MeV, as the channel $\pi\Omega^-$
strongly coupled to ${ \overline{ K} \Xi(1530)} $ opens. It seems that width of the isospin-1  ${ \overline{ K} \Xi(1530)} $ bound state is not compatible with the small
experimental width of $\Omega(2012)$.

As we discussed above in the molecular scenario, the $\Delta(1700)$ can not be the partner of $\Omega(2012)$ and in this case one should expect one additional excited $\Omega$
hyperon with $J^P=3/2^-$ in mass region of 2000-2150~MeV (see discussion in \cite{Guzey:2005vz}).

The simplest experimental way to figure out the nature of $\Omega(2012)$ (genuine resonance or ${ \overline{ K} \Xi(1530)}$ bound state) is to measure its branching for the $\Xi\overline{ K} \pi$ decay mode
and to search for its possible charge partners.

\noindent
\section*{\normalsize \bf Summary and outlook}

We have found that the recently detected, by the BELLE collaboration, $\Omega(2012)$  hyperon and the well known  $\Delta(1700)$ resonance can belong to the same 
$SU(3)$ decuplet of baryons with $J^P=3/2^-$. Mass splitting in the new decuplet is determined with good accuracy $\delta_{10}=104\pm 15$~MeV, which can be used as a benchmark
for various models of baryons.
Properties such as masses and decay widths of the still missing $\Sigma$ and $\Xi$ members of the decuplet are predicted in details (see Table~\ref{tab.:sigmaxi}), these predictions can guide experiments on hyperon spectroscopy.
 
 Our identification of the
decuplet is correct only if the sum of branching ratios for the decays $\Omega(2012)\to \Xi \overline{K}\pi, \Omega^-\pi\pi$ is not too large ($\le 70$\%).

For the large branching ratios of $\Omega(2012)\to \Xi \overline{K}\pi, \Omega^-\pi\pi$ decay modes, the most probable interpretation of the $\Omega$ hyperon is the $S$-wave $\overline{K} \Xi(1530)$
bound state
with quantum numbers also $J^P=3/2^-$. We note that the $\overline{K} \Xi(1530)$ bound  state can have isospin one and hence it has an additional decay mode  $\pi\Omega^- $.
 In a molecular picture, estimated width of $\Omega(2012)$ hyperon is $\sim 10$~MeV for isospin zero (decuplet), and considerably larger for the isospin one (27-plet).
 The latter option, seems,  is incompatible with the small
experimental width of $\Omega(2012)$.
A measurement of the three-body decay branchings of new $\Omega(2012)$ and search for the charge partners of it are crucial tests of  the nature of this hyperon.

We found also interesting that the studies of the strange hyperon properties can be a big help to constrain considerably  the (frequently large) uncertainties of the classical PWA of $\pi N$ processes.
It is not surprising -- mathematically PWA belongs to the class of so-called ill-posed (in Hadamard's sense \cite{JH}) problems and the $SU(3)$ relations  can provide a regularisation of the
ill-posed problem (see e.g. \cite{TA}). 

\noindent
\section*{\normalsize \bf Acknowledgements}

\noindent
MVP is grateful to K.~Semenov-Tian-Shansky for discussions of the effective chiral Lagrangian for baryons of arbitrary spin and to A. Gasparyan  and M.~Lutz for discussion of the Weinberg-Tomozawa term.
This work is supported by CRC110 (DFG).


\end{document}